# Characterization of the VEGA ASIC coupled to large area position-sensitive Silicon Drift Detectors

R. Campana,[ab,*] Y. Evangelista,[cd] F. Fuschino,[ab] M. Ahangarianabhari,[e]
D. Macera,[e] G. Bertuccio,[e] M. Grassi,[f] C. Labanti,[ab] M. Marisaldi,[ab] P. Malcovati,[f]
A. Rachevski,[g] G. Zampa,[g] N. Zampa,[g] L. Andreani,[bh] G. Baldazzi,[bh]
E. Del Monte,[cd] Y. Favre,[i] M. Feroci,[cd] F. Muleri,[cd] I. Rashevskaya,[g] A. Vacchi,[g]
F. Ficorella,[j] G. Giacomini,[j] A. Picciotto[j] and M. Zuffa[b]

[a] *INAF/IASF-Bologna, via Gobetti 101, I- 40129 Bologna, Italy*
[b] *INFN Bologna, Viale Berti Pichat 6/2, I-40127 Bologna, Italy*
[c] *INAF/IAPS, Via del Fosso del Cavaliere 100, I-00133 Roma, Italy*
[d] *INFN Roma Tor Vergata, Via della Ricerca Scientifica 1, I-00133 Roma, Italy*
[e] *Politecnico di Milano, Department of Electronics, Information and Bioengineering and INFN-Como Campus, Via Anzani 42, I-22100 Como, Italy*
[f] *University of Pavia, Department of Electronics, Information and Biomedical Engineering and INFN Pavia, Via Ferrata 3, I-27100 Pavia, Italy*
[g] *INFN Trieste, Padriciano 99, I-34149 Trieste, Italia*
[h] *University of Bologna, Dept. of Physics and Astronomy, Viale Berti-Pichat 6/2, I-40127 Bologna, Italy*
[i] *University of Geneva, DPNC, 24 quai E. Ansermet, Geneve, CH1211, Switzerland*
[j] *Fondazione Bruno Kessler, via Sommarive 18, I-38123 Trento, Italy*
*E-mail:* campana@iasfbo.inaf.it

ABSTRACT: Low-noise, position-sensitive Silicon Drift Detectors (SDDs) are particularly useful for experiments in which a good energy resolution combined with a large sensitive area is required, as in the case of X-ray astronomy space missions and medical applications. This paper presents the experimental characterization of VEGA, a custom Application Specific Integrated Circuit (ASIC) used as the front-end electronics for XDXL-2, a large-area (30.5 cm$^2$) SDD prototype. The ASICs were integrated on a specifically developed PCB hosting also the detector. Results on the ASIC noise performances, both stand-alone and bonded to the large area SDD, are presented and discussed.

---

[*] Corresponding author.

**Contents**

**1. Introduction**

In the field of X-ray spectroscopy, Silicon Drift Detectors (SDD) of small sensitive area, up to about 1 $cm^2$, are successfully exploited [1], being especially appreciated for the minimization of the electronic noise due to their small anode capacitance. However, some applications such as high-energy astrophysics instrumentation (X-ray all-sky monitoring, large-area X-ray timing experiments, gamma Compton cameras), and single-photon-emission computed tomography systems based on the Compton camera design for medical imaging, require the coverage of large sensitive areas. This makes the employment of the usual small-area detectors effectively impractical. Recently, monolithic SDDs with larger sensitive areas (up to several tens of $cm^2$) coupled to low noise integrated Front-End Electronics (FEE) have shown promising performance. High resolution spectroscopy not only depends on the quality of the SDD itself (the dark current can be minimized by cooling the detector), but also on the FEE performance. Therefore, to achieve high spectral resolutions a sophisticated design of the integrated front-end circuit is required.

In this paper we present and discuss the characterization of an integrated system composed of VEGA, a low-noise, low-power ASIC designed and optimized for analog pulse processing of signals from monolithic large area SDDs, coupled to XDXL-2, a 30.5 $cm^2$ SDD prototype.

The SDD is described in Section 2.1. The specifications of the ASIC are shown in Section 2.2 and the whole integrated system is described in Section 2.3. The experimental results are discussed in Section 3. Finally, Sections 4 and 5 report applications of the detector system and the conclusions, respectively.

## 2. The first integrated VEGA+SDD prototype

### 2.1 The Silicon Drift Detector (SDD)

Very large area multi-anode SDDs were developed for particle tracking applications such as the LHC-ALICE experiment at CERN [2]–[5], demonstrating very good performance in localizing the impact point of ionizing particles in a high multiplicity environment [6], [7]. Due to the low collecting capacitance and low current at each anode, these SDDs are well suited also for X-ray spectroscopy applications [8], [9]. Based on the large experience coming from the development of the ALICE SDD (by INFN Trieste, in collaboration with Canberra Industries Inc.) a new large area SDD optimized for X-ray applications was developed in the framework of a R&D activity carried on with a collaboration between INFN Trieste and Fondazione Bruno Kessler (FBK) in Trento, Italy. This work resulted in the production of various SDD prototypes, with different design choices [10]. The prototype used in the present case, XDXL-2, features a thickness of 450 μm and a total sensitive area of 30.5 cm$^2$ split in two detector halves. In order to study different possible applications (e.g. imaging or X-ray spectro-timing) the anodes placed at the edges of each side have different pitches, 147 μm and 967 μm, with capacitances of about 80 fF and 350 fF, for a total of 967 and 147 anodes, respectively. A negative HV potential is applied to the detector from its central cathode, while several integrated voltage dividers provide both the scaled potentials to create the constant drift field and the bias of the guard cathodes placed at the sides of the drift region. The latter are used to scale down the drift potentials to ground at the edge of the SDD in a controlled way. The X-ray photon interaction creates a cloud of electron–hole pairs. The electrons are quickly focused in the middle plane of the detector, thanks to the sideward depletion technique [11], and then carried towards the anodes thanks to the drift field. Because of diffusion, the size of the electron cloud increases with time, up to several hundreds of μm, before reaching the collecting anodes [9]. This phenomenon has important consequences on the energy resolution of the detector, degrading the spectroscopic performance with increasing drift lengths, when the electron cloud is read-out by more than one anode. However, in this work we only considered single-anode events (i.e. the events in which all the charges are collected by one anode), in order to estimate the noise performance of an ASIC channel reading out an anode of the large area device. For this reason, the noise measurements discussed in Section 3.2 were performed with the largest anode pitch (967 μm) only.

### 2.2 The VEGA ASIC

The VEGA ASIC, an integrated Front-End Electronics for large area linear SDDs, was designed and developed by a collaboration between Politecnico di Milano and University of Pavia. The VEGA ASIC consists of analog and digital/mixed-signal sections fulfilling the required low-noise, low-power specifications and functionalities for high-resolution X-ray spectroscopy in the energy range between 0.5 and 60 keV. In particular, the required energy resolution is ≤ 260 eV FWHM at 6 keV, which corresponds to an Equivalent Noise Charge (ENC) of the FEE/detector system ≤ 27 e$^-$ RMS. Furthermore, the ASIC is designed to be compliant with the integration of an embedded ADC, not yet included in this design. The FEE circuit was designed to have less than 500 μW total power consumption per channel. The low power consumption makes the VEGA ASIC especially suitable for application in systems where a large number of detector channels must be read-out with limited resources in terms of power budget. The ASIC has been designed and manufactured with Austriamicrosystems 0.35 μm CMOS C35B4C3 technology [12]. The ASIC design includes a single-cell version, for testing purposes, and a 32 channel array with a dimension of 200 μm × 500 μm per channel.

The analog section of each channel includes a low-noise charge-sensitive preamplifier (CPA), a CR-RC shaper and a peak stretcher/sample-and-hold circuit. The nominal shaping times are selectable in the range from 1.6 to 6.6 μs. The digital section is composed of an amplitude discriminator, with a threshold that can be set by both coarse and fine levels, and a peak discriminator. Moreover, pile-up rejection, signal sampling, generation of the trigger, the reset of the channels and configuration of the ASIC are performed by a dedicated logic section. The ASIC is configured by sending a 247-bit word on the three relevant configuration pins (Enable_Writing, CLK, Data_Serial_In). The configuration includes global and single channel threshold setting, enabling/disabling of each individual channel preamplifier and discriminator, shaping time setting, bias voltage of the input stage feedback pMOS, shaped/stretched signal output selection, and trigger mode selection. In particular, the trigger logic can be set to activate the signal stretching independently or simultaneously with the first triggering channel. The shaped or stretched signal for each channel is multiplexed at the ASIC output. Finally, the ASIC is equipped also with a selectable test circuit, with an input test capacitance of 20 fF, which allows the injection of an external charge to the preamplifiers.

In Figure 1 a block diagram of the ASIC is presented, while full details on the design and the functionalities of the VEGA ASIC are presented in [13] and references therein.

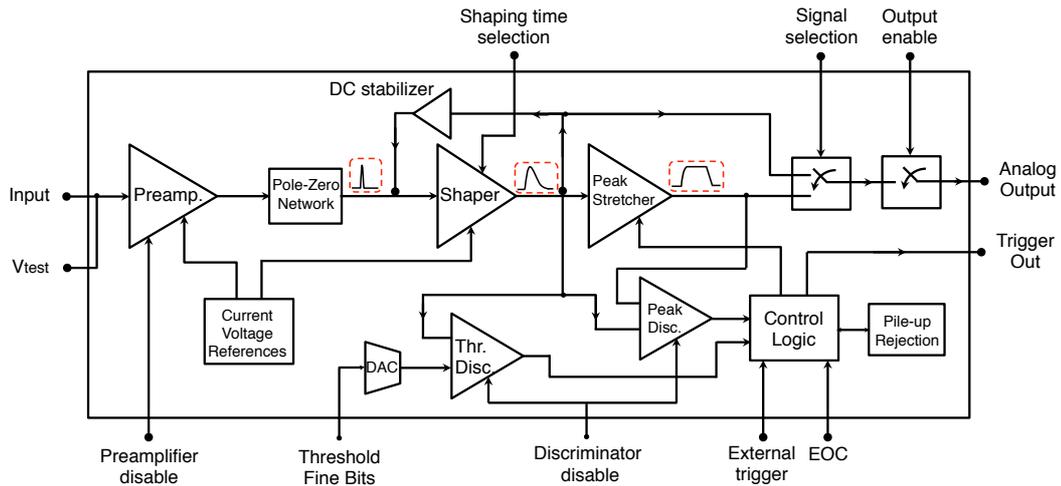

**Figure 1** - Functional block diagram of the VEGA ASIC.

## 2.3 The integrated system

In order to characterize the multichannel VEGA ASIC connected with the large area SDD, an integrated test system was realized, combining an electronic test board, a power supply system and an acquisition equipment.

The electronic test board, designed and realized by the University of Geneva, handles both analog and digital signals. The printed circuit board (PCB) is organized to accommodate the large area SDD (Section 2.1), four VEGA ASICs (Section 2.2) and the electronic components to interface them with the test equipment. Power supplies for the analog and digital sections of the ASICs are derived by means of low-dropout (LDO) regulators and filters from common ±5 V input voltages.

The four ASICs are mounted in pairs on each side of the SDD. The two ASICs belonging to the same SDD side, i.e. with the same anode pitch, are read-out simultaneously. Figure 1 shows the whole integrated system with the large area SDD and four ASICs. The 68-pin connectors on both sides of the board provide the interface with the test equipment.

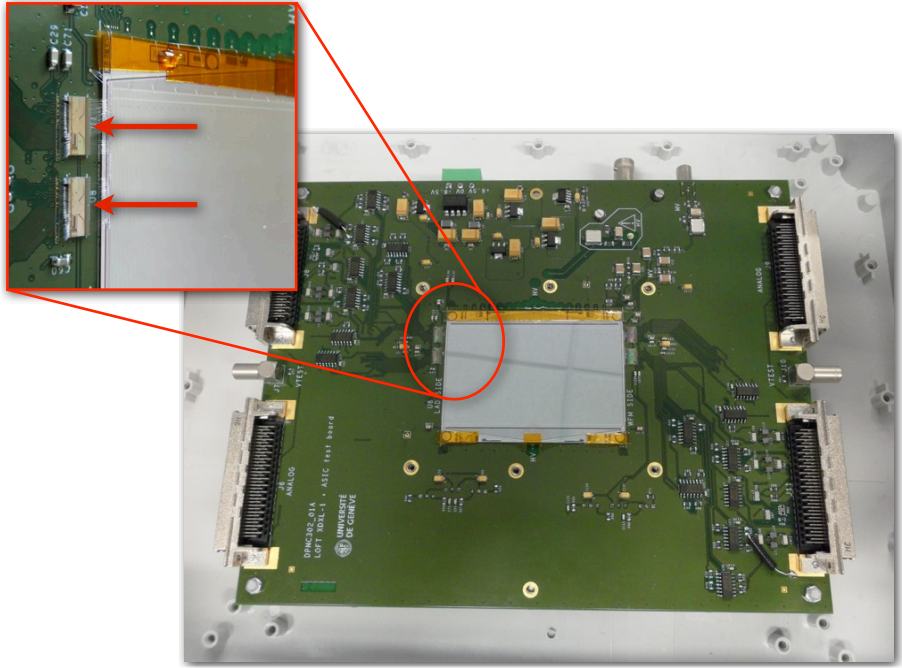

**Figure 2 -** The integrated setup used in this work. Both the SDD and the four VEGA ASICs (in the upper left and right corners of the SDD) are visible. The connectors on both sides provide the interfaces to the test equipment. The ASICs are shown also in the inset, indicated by the red arrows.

The PCB is housed in an aluminum box in order to shield it from electromagnetic interferences. The box was designed also to allow the use of Peltier cells and water-cooled heat sinks for the use of the system outside the laboratory, in which we instead used a climatic chamber for the characterization at different temperatures.

The test equipment is based on a PC equipped with two National Instruments boards and a custom-made software implemented using the LabVIEW environment. The first board, NI-7811R, is a Multifunction RIO Device, equipped with 160 digital input/outputs and a Virtex-II XC2V1000 FPGA. This board is used to set the configuration of the two ASICs under test and to manage the read-out digital signals. The second board, a NI-6133, is a Simultaneous Multifunction Input/Output DAQ system providing 8 differential analog inputs and a 14-bit ADC operable up to 2.5 MS/s per channel. This board is used for data acquisition. Figure 3 shows the configuration panel for a single ASIC, where both the individual channel settings and the settings common to all channels are visible. Figure 4 shows the data acquisition panel.

The stretched analog output signals of the 32 channels of each ASIC are multiplexed at the ASIC output and read out sequentially. The test equipment software then samples the multiplexer waveform and stores the ADC value of each channel.

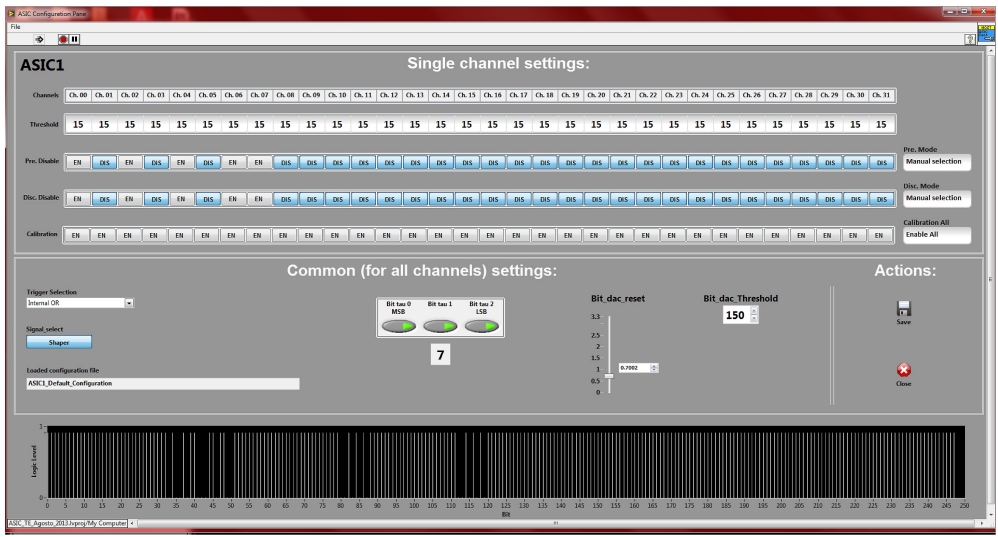

**Figure 3 -** Screenshot of the test equipment LabVIEW panel for the ASIC configuration. Both single channel settings (upper part) and common settings (lower part) are visible.

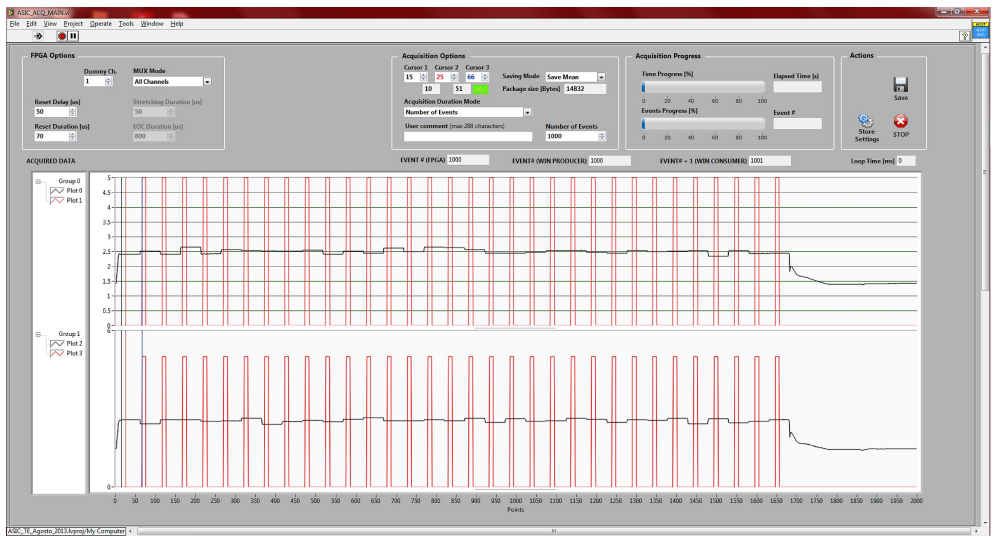

**Figure 4 -** Screenshot of the LabVIEW panel for the data acquisition of both ASICs. The upper black curve represents the stretched signal of all 32 channels of the ASIC #1 multiplexed at the chip output when all ASIC channels are stimulated with an input pulse of 100 mV (equivalent to ~45 keV in silicon). The lower black curve represents the output signal of the ASIC #2 with the same stimuli. The first multiplexed acquisition of each chip is rejected due to the particular implementation of the data storage. The multiplexer cycles between the various channels each 20 μs. The red curve show the intervals in which the ADC sampling is made.

In order to characterize the contribution to the overall noise of both the ASIC itself and the SDD, experimental measurements were carried out before and after the SDD integration with the ASICs. As a first step (Section 3.1), the ASICs were tested alone, using a spectroscopic pulser (a BNC DB-2 NIM model) as the test input source, estimating the gain and noise for each channel. After the SDD-ASIC integration, we performed tests with radioactive sources ($^{55}$Fe and $^{241}$Am) in the climatic chamber of the IAPS-Rome X-ray facility, evaluating the full performance of the system.

A non-negligible common mode noise (CMN, i.e. a noise component present on all anodes and fully correlated to a common noise source) has been observed, originating within the detector bias, and coupled to the anodes through their capacitances with the nearby electrodes. The CMN was already reduced by decoupling these cathodes directly on the PCB hosting the detector and filtering the power supplies. A better performance can however be obtained by removing the residual common noise from the data with a direct estimation. This analysis method, described in detail in [8], requires that all channel signals are held and acquired synchronously with the first triggering channel. The CMN can then be removed by estimating the event-by-event baseline shift using the channels where no signal charge was integrated. The final result of the common mode noise correction is strongly dependent on the number of channels involved in its estimation. Figure 5 reports the *residual* contribution to the overall noise due to this CMN correction as a function of the total number of channels used to evaluate the event baseline. For this simulation, all the ASIC channels are assumed to have the same electronic noise charge, equal to 17 e$^-$ RMS. In this case, the residual noise due to the CMN subtraction is simply $\sigma^2_{res} = \sigma^2/M$, where $\sigma^2$ is the channel ENC and M is the total number of channels used to evaluate the baseline [8].

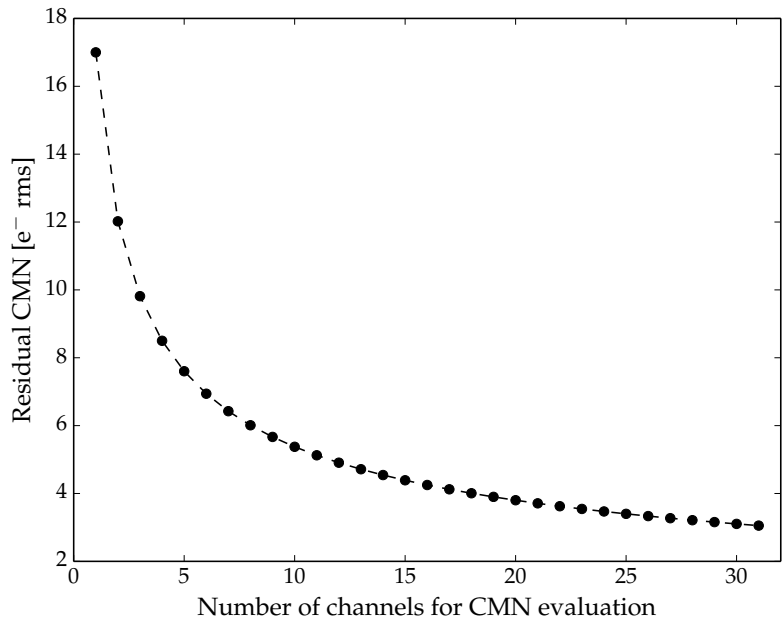

**Figure 5 -** Simulation of the residual contribution to the total energy resolution from the common-mode noise subtraction, as a function of the number of channels used to evaluate it. An electronic noise of 17 e- RMS is here assumed for all channels.

**2.4 Pulser measurements**

To characterize the noise performance of the ASICs, three different test pulse amplitudes of 20 mV, 50 mV and 100 mV were used, corresponding to 2500 e$^-$, 6250 e$^-$ and 12500 e$^-$ respectively (equivalent to 9 keV, 22 keV and 45 keV signals in silicon). Figure 6 shows the distribution of the gain measured simultaneously on 62 channels (31 for each ASIC: the acquisition of the last channel of each ASIC was unavailable due to a test equipment issue). A gain variation of about 5% RMS with respect to the average value is evident, which is compatible with the fabrication process tolerance.

Figure 7 shows the equivalent noise charge distribution, in e$^-$ RMS, measured simultaneously on all channels. The measurements were done estimating the FWHM on the 20 mV amplitude peaks. The average value of the noise over the channels is 14.1 e$^-$ RMS. The dispersion in the noise values is about 0.7 e$^-$, with only one channel significantly noisier than the average.

As already verified during testing of the single-cell ASIC version [13], the gain is linear, with a linearity error better than 1% in the equivalent energy range explored.

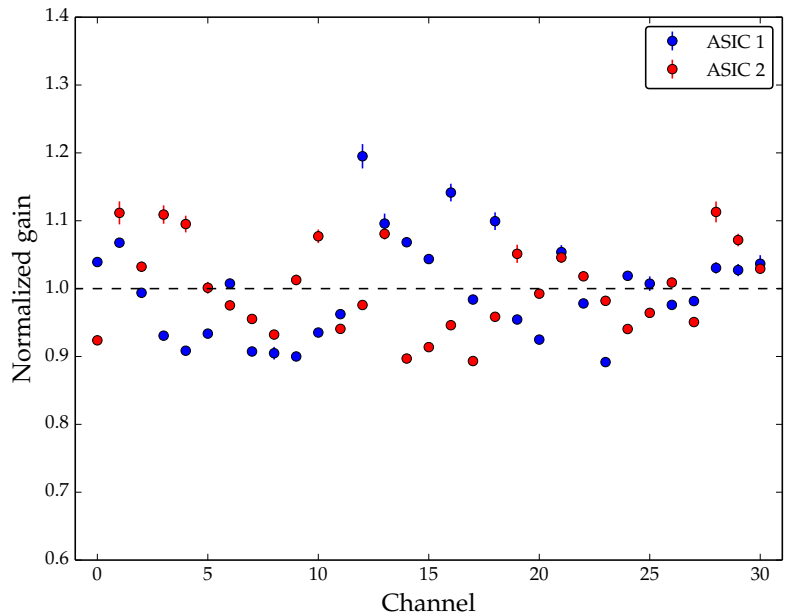

**Figure 6 -** Gain distribution measured on the available channels of two different ASICs (31 channels for each ASIC). The data points have been normalized with respect to the mean value. The maximum variation in the gain distribution is within 20% of the average value.

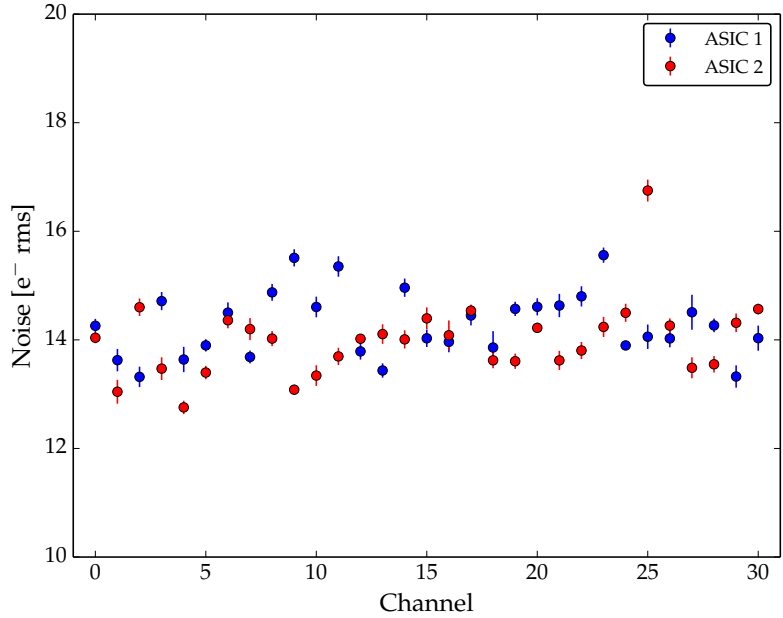

**Figure 7** - Noise distribution measured on the available channels of two different ASICs. The average value of the noise over the channels is 14.1 e⁻ RMS.

## 2.5 X-ray measurements

After the characterization of the stand-alone ASICs, the SDD was integrated on the PCB and a X-ray characterization campaign was performed at the INAF/IAPS X-ray facility [14]. The integrated detector (Figure 8, see also the inset of Figure 1) was placed in a climatic chamber where non-collimated $^{55}$Fe spectra were acquired at a detector temperature of -30 °C, where the leakage current is ~4 pA per channel. The aluminum box containing the detector was continuously exposed to a nitrogen flow in order to maintain a dry atmosphere.

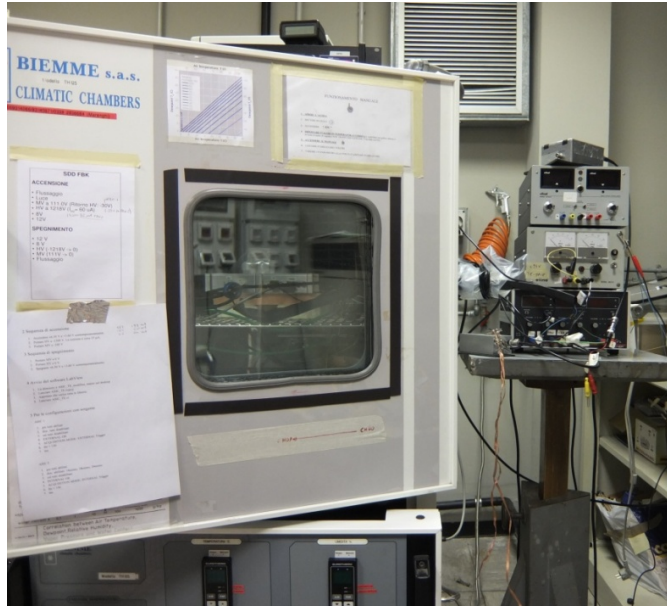

**Figure 8** - Integrated system in the INAF/IAPS climatic chamber

The data were analyzed rejecting events where the signal charge was collected by more than one anode in order to accumulate the spectra for single-anode events only. Since only 8 or 9 channel per ASIC were bonded (due to the mismatch between the 967 μm anode pitch and the 200 μm ASIC channel pitch), after excluding the neighboring channels to the triggered one only five channels are available for common noise estimation.

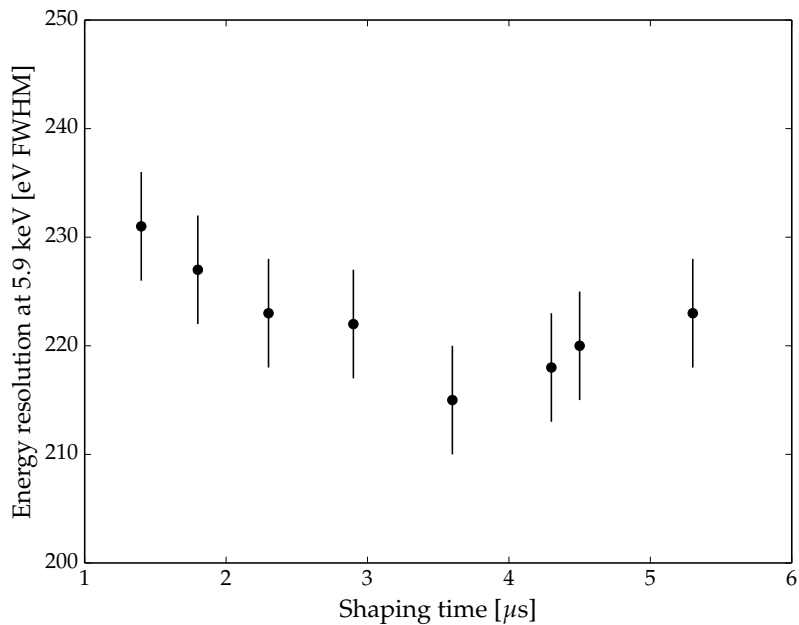

**Figure 9** - Energy resolution (expressed as the FWHM of the $^{55}$Fe line at 5.9 keV) as a function of the measured shaping time.

Figure 9 shows the energy resolution (after the CMN subtraction) obtained for one ASIC channel on the Mn Kα line as a function of the shaping time. The measured shaping times resulted to be about a factor 0.77 shorter than the nominal ones. This behavior is due to fabrication process tolerances. As it can be seen in Figure 9, a shaping time of 3.6 µs was determined to be optimal at a temperature of -30° C. Error bars in Figure 9 were estimated by comparing different data sets acquired with the same ASIC configuration and experimental conditions. A systematic maximum dispersion of ~10 eV was found, which is much larger than the expected statistical uncertainty (of the order of 1.5 eV RMS) and points out the presence of a time varying noise component not completely rejected by the power supply filtering circuits. Further optimizations of the ASIC functional controls (analog chain, I/V references, input stage of FEE circuit) were carried out in order to minimize the system noise, thus further enhancing the instrument spectral resolution.

In Figure 10 and Figure 11 we show the $^{55}$Fe energy spectrum acquired in the optimized condition, before (Figure 10) and after (Figure 11) the CMN correction, for a single channel. As it can be seen, an energy resolution of 288 eV FWHM at the 5.9 keV Mn Kα line is obtained without the CMN correction, while an energy resolution of 205 eV FWHM (corresponding to an equivalent noise charge of 19.8 e$^-$ RMS for the complete detector-ASIC system) is reported after the correction. However, this value is affected by the incomplete subtraction of the CMN due to the low number of channels (see also Figure 4). Taking this into account, the intrinsic electronic noise can be estimated to be ~18.1 e$^-$ RMS.

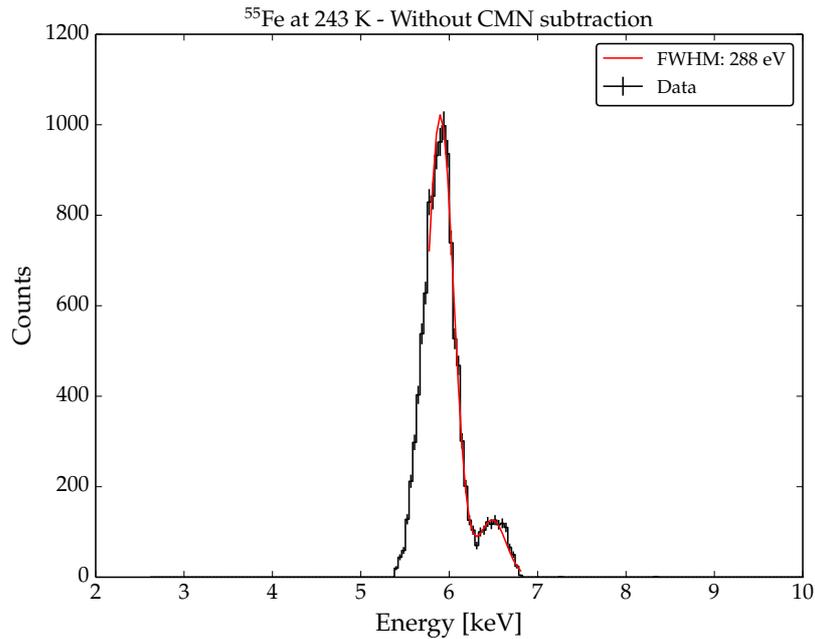

**Figure 10** - $^{55}$Fe spectrum taken at -30 °C (~4 pA/channel) using a shaping time of 3.6 µs. Single-anode events were selected rejecting events that share signal charge with neighboring anodes. No correction for the common-mode noise has been performed in this case. The energy resolution is 288 eV FWHM for the Mn Kα line (31.1 e$^-$ RMS).

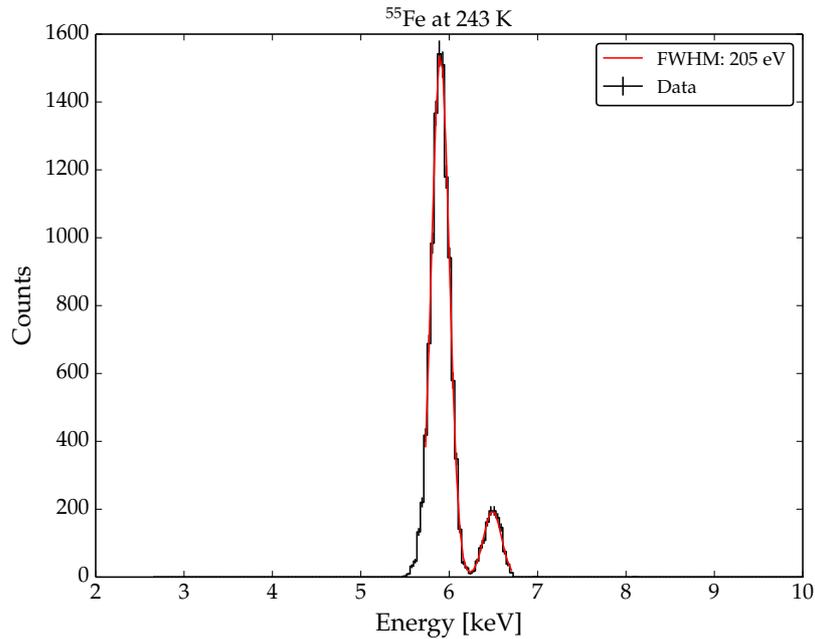

**Figure 11** - ⁵⁵Fe spectrum taken at -30 °C (~4 pA/channel) using a shaping time of 3.6 μs. Single-anode events were selected rejecting events that share signal charge with neighboring anodes. The energy resolution is 205 eV FWHM for the Mn Kα line (19.8 e⁻ RMS).

## 3. Applications

Large-area multi-anode SDDs coupled with low-noise, low-power integrated FEE make them extremely attractive for space astrophysics and medical applications. Here, low resources in terms of weight, volume, and power, combined with good energy and position resolution, and large sensitive area are usually required.

In the following sections a brief description of example space and medical applications is presented.

### 3.1 The LOFT ESA Mission

Large-area SDDs with low-noise, low-power ASIC readout are the fundamental keys to LOFT, the Large Observatory for X-ray Timing [15], a medium-class space mission selected by the European Space Agency (ESA) in February 2011 for a 3-year assessment phase. The LOFT project was proposed within the context of the Cosmic Vision Program (2015-2025). The LOFT payload consists of a Large Area Detector (LAD) [16], and a Wide Field Monitor (WFM) [17], both based on the large-area SDDs. An up-to-date table summarizing the anticipated performances of the LAD and WFM is provided on the mission web-site [18]. LOFT is specifically designed to investigate "matter under extreme conditions". By taking advantage of the large collecting area and fine spectral resolution of its instruments, LOFT will exploit the technique of X-ray timing combined with spectroscopic measurements.

The LAD instrument is a collimated non-imaging instrument performing pointed X-ray observations, mainly in the 2-30 keV energy band. It comprises six panels organized in 21 modules with 16 SDDs (sensitive area of ~76 cm²) per module. A total of 2016 detectors (i.e., 126 modules) are needed to complete the whole LAD, resulting in a total effective area > 10 m$^2$ at ~10 keV (geometric area of ~18 m$^2$). The LAD SDDs are equipped with lead-glass collimators [19] to limit the instrument field of view (FOV) to within ~1 degree up to energies of ~30-40 keV.

The WFM detector is a coded-mask instrument whose main goal is to scan large fractions of the sky simultaneously to catch interesting targets that can be followed up by the LAD. The WFM comprises five units composed of ten single direction cameras orthogonally arranged. Each camera is equipped with its own coded-mask and four SDDs similar to those employed for the LAD (optimized for imaging purposes). The main operating energy range of the WFM is 2-50 keV. The ten cameras of the WFM together achieve a FOV comprising roughly a rectangular region of 180°×90°. The on-board software of the WFM (LOFT Burst alert System) is able to identify and localize bright transient events with ~arcmin accuracy.

The ASIC system described in this paper is particularly suitable, from a performance standpoint, for the readout of the SDDs of both instruments. The LAD required energy resolutions are 260 and 200 eV FWHM for double and single-anode events, respectively, corresponding to a channel ENC of 19 e$^-$ RMS.

### 3.2 Medical diagnostics

In medical applications the use of SDDs is suggested for a Compton camera configuration, where the position of the primary interaction is determined by a tracker made of SDD detectors. Compton cameras are used to reconstruct gamma-ray emitting radioisotope distributions: the choice of detector material, the pixel size and the camera geometry all contribute to the overall performance. The use of a tracker made of large-area SDD detectors and ASICs provide for an optimal angular resolution and efficiency of the camera. Furthermore, for these applications the SDD electronics can be triggered by other (fast) detectors, allowing for a full 2D position reconstruction capability.

SDDs can also be used in radiology as detectors for X-ray monochromatic beams. A single beam composed of photons with three different energies invests the patient, being absorbed in a different way by the various tissues. The SDD, thanks to its high energy resolution performance, can be used in order to discriminate the different energies. In other uses the SDD may also be coated with a scintillator. The range of applications of such an instrument is vast and reaches beyond those of diagnosis in nuclear medicine: monitoring and decommissioning of nuclear power plants, homeland security, and materials study are just few example where a large area SDD can be profitably exploited.

### 4. Conclusions

The low temperature (-30 °C) spectroscopy performance of a 30.5 cm$^2$ Silicon Drift Detector integrated with low-power and low-noise front-end electronics have been presented. This is a prototype demonstrator that can be profitably used for experiments and applications requiring good spectroscopy resolutions, 205 eV FWHM at 6 keV, combined with a very-large sensitive area. The key point of the integrated system performance relies on the 32-channel front-end VEGA ASIC prototype. Work is ongoing to realize an SDD with a sensitive area of ~76 cm$^2$, which represents the proposed baseline for the LOFT mission.

**Acknowledgments**

The results described in this paper were obtained within the INFN projects XDXL (X-ray Drift eXtra Large) and ReDSoX (Research Detectors for Soft X-rays (ReDSoX). The work was also partially supported under ASI agreement I/021/12/0. We are grateful to Federico Ferrari for his help in developing the test equipment software.